# HEW simulations and quantification of the microroughness requirements for X-ray telescopes by means of numerical and analytical methods

D. Spiga[a*], G. Cusumano[b], G. Pareschi[a]

[a]INAF/ Osservatorio Astronomico di Brera, Via E. Bianchi 46, I-23807 Merate (LC), Italy
[b]INAF/Istituto di Astrofisica Spaziale e Fisica Cosmica, Via La Malfa 153, I-90146 Palermo, Italy

**ABSTRACT**

Future X-ray telescopes like SIMBOL-X will operate in a wide band of the X-ray spectrum (from 0.1 to 80 keV); these telescopes will extend the optical performances of the existing soft X-ray telescopes to the hard X-ray band, and in particular they will be characterized by a angular resolution (conveniently expressed in terms of HEW, Half-Energy-Width) less than 20 arcsec. However, it is well known that the microroughness of the reflecting surfaces of the optics causes the scattering of X-rays. As a consequence, the imaging quality can be severely degraded. Moreover, the X-ray scattering can be the dominant problem in hard X-rays because its relevance is an increasing function of the photon energy. In this work we consistently apply a numerical method and an analytical one to evaluate the X-ray scattering impact on the HEW of an X-ray optic, as a function of the photon energy: both methods can also include the effects of figure errors in determining the final HEW. A comparison of the results obtained with the two methods for the particular case of the SIMBOL-X X-ray telescope will be presented.

**Keywords:** X-ray telescopes, Half-Energy-Width, X-ray Scattering, Power Spectral Density

## 1. INTRODUCTION

The launch of a number of X-ray observatories, in addition to those already operating (Chandra, XMM-Newton, SWIFT), is foreseen in the near future. Some of them, like SIMBOL-X[1], Constellation-X/HXT[2], NeXT[3], XEUS/HXT[4], will extend the current techniques for X-ray focusing (to date, adopted only below 10 keV) to the hard (E > 10 keV) X-ray band by adopting very shallow incidence angles (0.1 ÷ 0.4 deg) and graded multilayer coatings[5,6]. For example, the SIMBOL-X sensitivity band is 0.5 ÷ 80 keV, with a required angular resolution of 15 ÷ 20 arcsec. Other X-ray observatories will instead take the advantage of a very large Field of View (1.5 deg diameter for the Wide Field Imager instrument aboard EDGE[7]) in a more limited energy band (0.3 ÷ 6 keV), made possible by the adoption of optics constituted by mirrors with polynomial profiles[8,9]. These mirror profiles enable, at the expense of a slight degradation of the on-axis angular resolution, a much better off-axis imaging quality with respect to the widespread Wolter-I mirrors,.

An important point for all these instruments is the request of a good angular resolution, heretofore expressed in terms of HEW (*Half Energy Width*). The optics, in particular, should be characterized by imaging performances comparable to those of XMM-Newton (15 arcsec HEW) and SWIFT-XRT (20 arcsec HEW). For this reason, the factors that could lead to a degradation of the HEW have to be carefully analyzed and corrected.

Departures of the mirror profile from the nominal one ("figure errors"), and mirrors misalignments, are a frequent factor of imaging degradation in X-ray optics. As far as the optical path deviations introduced by these defects are much larger than the photon wavelength, λ, the focal spot blurring can be treated with geometrical optics tools. Therefore, the impact of the deformations on the HEW - considered as independent of λ - can be evaluated along with ray-tracing codes.

---

* e-mail: daniele.spiga@brera.inaf.it, phone +390399991146, fax +39039999160

Another breakthrough, however, can occur as a consequence of reflecting surfaces *microroughness*. When X-rays impinge at a grazing-incidence angle $\theta_i$ on a rough surface with rms σ, its reflectivity decays exponentially, following the well-known *Debye-Waller formula*[5]:

$$R_\sigma = R_0 \exp\left(-\frac{16\pi^2 \sin^2 \vartheta_i \sigma^2}{\lambda^2}\right),\qquad(1)$$

where $R_\sigma$ is the measured reflectivity in the direction specular to that of incidence and $R_0$ is the reflectivity as computed from the Fresnel equations (i.e., for an ideally smooth mirror). Besides the reflectivity reduction, the missing X-ray photons are *scattered* in the surrounding directions. X-ray Scattering (XRS) from rough surfaces is a well-known and studied effect[10,11] that can seriously degrade the angular resolution, because it is an increasing function of the photon energy *E*. Therefore, the XRS contribution to the HEW is of increasing relevance as one moves to the hard band of the X-ray spectrum, where it can even dominate the figure error term.

XRS is a surface diffraction effect, taking place whenever the optical path dispersion due to the surface relief becomes comparable to λ. For this reason, the limit between figure errors and XRS is not fixed but depends on $\theta_i$ and λ. A possible criterion (proposed by Aschenbach[12]) considers the rms of each Fourier components of the surface. If this component fulfills the smooth-surface limit, i.e., $4\pi\sigma \sin\theta_i < \lambda$, is dominated by roughness; otherwise, it should be mainly considered as geometric deformation. We plot in Fig. 1 the limiting σ values that mark the boundary between the two regimes, in the energy and incidence angles range of interest for X-ray telescopes. Considering the typical relations between incidence angles and photon energies in X-ray optics, the boundary on the rms for a single spatial frequency lies between 0.5 and 2 nm for the telescopes SIMBOL-X, EDGE and XEUS.

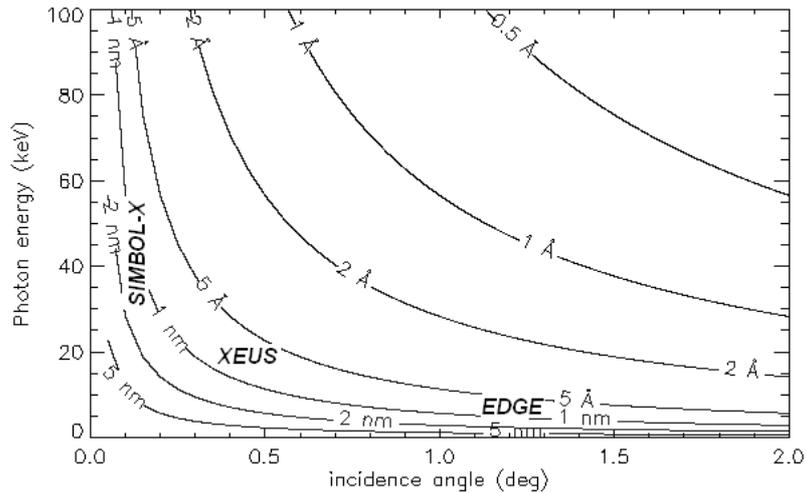

Fig. 1: separation of the mirror figure error and XRS to the HEW. Angles and energy ranges for the SIMBOL-X, EDGE and XEUS X-ray telescopes are also indicated. We did not highlight the full energy ranges of the telescopes to improve the figure readability.

In order to keep the HEW of X-ray optics for imaging telescopes at the required level for a scientific relevance of the images, clear requirements on the mirror figure and microroughness have to be stated. In the following we shall focus on the problem of defining surface roughness tolerances, by means of the simulation of the expected HEW of an X-ray mirror module. To this end, the surface roughness will be conveniently expressed in terms of its *Power Spectral Density*[11] (PSD), because it is the physical quantity that directly determines the intensity and the distribution of the XRS. The PSD provides us with a detailed characterization of the surface roughness amplitude as a function of the spatial frequency *f* (or, equivalently, the spatial wavelength $\ell = 1/f$). The rms roughness σ in a given spectral band $\Delta f$ can be obtained by integration of the PSD *P(f)*:

$$\sigma^2 = \int_{\Delta f} P(f)\,df.\qquad(2)$$

The σ parameter will therefore depend on the particular spectral band of interest or sensitivity. A frequent PSD modelization for an optically-polished surface is a *power-law spectrum*[11, 12]:

$$P(f) = \frac{K_n}{f^n}, \qquad (3)$$

where *n* is a real number taking on values in the interval (1: 3) and $K_n$ is a normalization constant. The roughness power spectrum of several substrates for X-ray mirrors well approximates this formula[14] in a wide spatial frequency range. X-ray mirrors, indeed, can deviate from this model, especially if a reflective coating has been deposited onto the substrate, owing to the roughness growth[15] that may occur – in a variable extent – in the high-frequency range. It should be noted that the σ parameter is not sufficient to describe the surface aspect or to return quantitative information of the HEW: surfaces with the same rms can be very different (see Fig. 2) and the same σ in different frequency ranges can have a completely different impact on the imaging degradation. For instance, given a fixed σ value, a smoothly-decreasing PSD (*n*~1) will comprise more high-frequency roughness than a steeply-decreasing one (*n*~3), and since high frequencies are responsible for large-angle XRS, it will cause a faster increase of the HEW with the photon energy. This is a simple consequence of the grating formula[10], that relates the scattering/incidence angles ($\theta_s$ and $\theta_i$) to *f*:

$$f = \frac{\cos\vartheta_i - \cos\vartheta_s}{\lambda}. \qquad (4)$$

The evaluation of the HEW from a surface PSD has been performed in the past years by several authors (e.g., De Korte et al.[16]; Christensen et al.[17]; Harvey et al.[18]; O'Dell et al.[19]; Willingale[20]; Zhao and Van Speybroeck[21]) using different methods to address the computation of the X-ray scattering distribution, to be convolved with the figure errors to return the mirrors PSF, and consequently the HEW. These methods, though very accurate, require the calculation of the whole PSF for a single photon energy, and thereby require a considerable computational effort. Hence, it would be advantageous to find out a proper methodology to infer the HEW from a direct calculation on the characteristic parameters of the surface roughness.

In this work we attempt to provide with quantitative predictions of the HEW for the SIMBOL-X optical module, as a function of the photon energy, starting from given assumptions on the surface finishing level. The most innovative feature of the SIMBOL-X[1] X-ray telescope will be the unprecedented extension of imaging capabilities to the hard X-ray band *up to 80 keV*. Even if this extension will be mainly made possible by the adoption of graded multilayer coatings, the incidence angles will have to be very shallow (see Tab. 1) to ensure a good reflectivity over the photon energy band of sensitivity. As the mirror module diameter is similar in size to XMM (700 mm), the focal length will be very long (20 m). With the present technology, such a focal length cannot be managed with a single spacecraft, therefore the *formation-flight* configuration[22] has to be adopted. Since the energy band of SIMBOL-X is extended up to the hard X-ray energy band, the mirrors surface smoothness has to be much better than for the case of XMM-Newton (σ ~ 7 Å) to fulfill the 20 arcsec HEW requirement at 30 keV (see Tab. 1), which is needed to resolve astronomical targets like the hard X-ray background (XRB) in the peak region. For this reason, the quantification of the HEW due to X-ray scattering is a very important issue in order to establish the surface smoothness tolerance to be required for the SIMBOL-X mirrors.

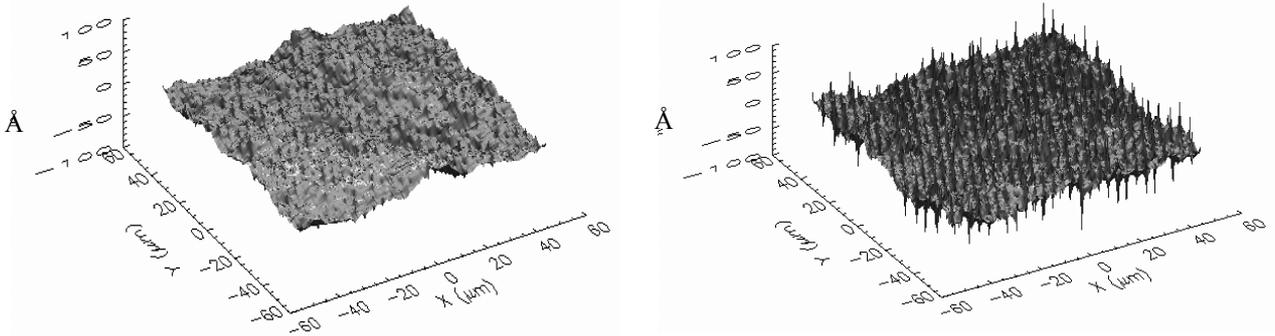

Fig. 2: two different *simulated* surfaces with the same size (100 μm) and rms (10 Å), but PSD with different spectral indexes: 2.3 (left) and 1.4 (right). The larger high-frequency content is apparent in the second case. Therefore, the imaging degradation will be more severe at high energies.

Tab. 1. Main features of the SIMBOL-X optical module, compared to those of XMM-Newton

|                          | XMM-Newton      | SIMBOL-X        |
|--------------------------|-----------------|-----------------|
| **Focal length**         | 7.5 m           | 20 m            |
| **Min diameter**         | 300 mm          | 300 mm          |
| **Max diameter**         | 700 mm          | 700 mm          |
| **Min incidence angle**  | 0.28 deg        | 0.11 deg        |
| **Max incidence angle**  | 0.67 deg        | 0.25 deg        |
| **Number of shells**     | 58              | 100             |
| **Energy band**          | 0.1 – 10 keV    | 0.5 – 80 keV    |
| **Effective area (1 keV)** | ~ 1400 cm²    | ~ 1400 cm²      |
| **Effect. area (30 keV)** | -              | ~ 450 cm²       |
| **Required HEW (1 keV)** | 15 arcsec       | 15 arcsec       |
| **Required HEW (30 keV)**| -               | 20 arcsec       |

To this end, we shall either make use of a numerical routine or follow an analytical approach. The numerical code (Sect. 2), based on the treatment discussed by Green et al.[23], was written by Sacco et al. (1996) and integrated in a raytracing FORTRAN program to interpret the PSF of the Beppo-SAX X-ray telescope on the basis of simple assumptions for the surface roughness of the X-ray mirrors. The analytical approach (Sect. 3) has been developed by one of us (D. Spiga[24]), and enables an immediate translation of a surface PSD into the expected XRS term of the HEW. In addition, it enables the reverse computation for a single mirror shell (from the HEW trend to the PSD). In Sect. 4 we deal with a comparison of an application of the two methods to the SIMBOL-X optical module. The results are briefly summarized in Sect. 5.

## 2. COMPUTATION OF THE SIMBOL-X HEW: MODIFIED RAY-TRACING ROUTINE

The simulation of the imaging degradation caused by mirrors deformations can be often performed along with *X-ray tracing routines*, simulating a set of rays that follow the geometrical optics laws. The same approach cannot be used to account for XRS, because the concept of ray becomes no longer applicable. However, the first-order approximation XRS theory[10] can be applied to correct the enlargement of the HEW obtained from usual X-ray routines as follows:

1. given a proper PSD (or, equivalently, a self-correlation function) modelization, the scattering intensity for X-ray photons (impinging on the optic at the incidence angle $\theta_i$) is calculated at each scattering angle $\theta_s$.
2. each photon used in the ray-tracing simulation is assigned a scattering likelihood proportional to the intensity of the X-ray scattering distribution.

The adopted procedure[23], rather than a PSD, takes as input two characteristic parameters of the surface roughness, that can be derived from the PSD. The *rms roughness* $\sigma$ (Eq. 2) and the *surface correlation length* $\tau$, that can be defined as the average spatial wavelength of the surface micro-relief. The parameter $\tau$ in a spatial frequency window $\Delta f$ can be computed along with the formula

$$\tau_{\Delta f} = \frac{2\pi \sigma_{\Delta f}}{m_{\Delta f}}, \tag{5}$$

where $m_{\Delta f}$ is the *rms slope*, measured in radians. This parameter is an index of the steepness of the rough features of the surface, and can be calculated from the PSD in the spectral band $\Delta f$:

$$m_{\Delta f}^2 = \int_{\Delta f} (2\pi f)^2 P(f) df \cdot \tag{6}$$

We can expect that the HEW, at a given photon energy, is an increasing function of the $\sigma$ parameter. Similarly, as a smaller $\tau$ indicates that the roughness is mainly concentrated at high frequencies (that scatter X-rays at higher angles), the HEW is expected to increase as $\tau$ decreases.

The numerical routine has been applied to 5 couples of $\sigma$ and $\tau$ parameters ($\sigma$ in the interval 3-4 Å, $\tau$ between 11 and 20 μm) for the entire SIMBOL-X module. The HEW has been directly derived from the simulated photon distribution on the focal plane, assumed to have a 310 arcsec radius, similar to that of SIMBOL-X. The HEW trends, from 1 to 60 keV, are plotted in Fig. 3. To account for possible mirror profile deformations, 15 arcsec figure error were added in quadrature to the calculated X-ray scattering terms.

The calculated trends exhibit a clear increase of the HEW with the photon energy, as expected. At a fixed photon energy, the HEW is an increasing function of the rms roughness, and is particularly sensitive to small variations of the correlation length. These results are in agreement with our qualitative discussion.

It is worth noting that the applied program is currently being used also to quantify the impact of the stray light in the SIMBOL-X optics, that also strongly depend on XRS caused by surface roughness. The evaluation of this effect, as well as the optimization of methods to reduce it, is also described in this volume[25].

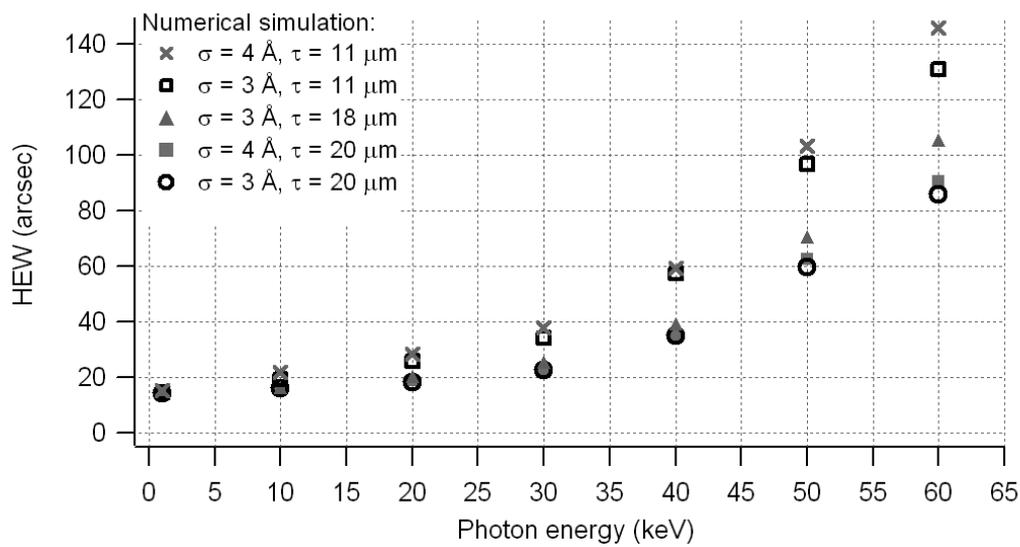

Fig. 3: results of the application of the numerical routine to the SIMBOL-X optical module, for different σ and τ values.

## 3. ANALYTICAL DERIVATION OF THE HEW FROM THE PSD

Another possibility to predict the optical performances of an X-ray optical module is the analytical translation of a surface PSD into the expected HEW[24]. This method, based on the well-known theory of X-ray scattering from rough surfaces[10], can be applied to any PSD, and holds for an optical system with an arbitrary number $N$ of reflections at the same *grazing* incidence angle $\theta_i$. Moreover, it is reversible: that is, from a given HEW($\lambda$) requirement one can derive a corresponding mirror surface PSD function, that can be assumed as *microroughness tolerance* for the mirrors surface.

We consider firstly a single mirror. The application of this method requires that:

1) X-ray reflection and scattering occur in grazing incidence, so that the XRS chiefly lies in the plane of incidence.
2) The mirrors surface is smooth, so that the XRS theory[10] can be applied, and the scattering angles are always small when compared with the incidence angle.
3) The surface is coated with either a single layer coating, or a graded multilayer with a smoothly decreasing reflectivity with the incidence angle and the photon energy.
4) The reflective coating roughness is described by a single PSD (i.e., no roughness growth throughout the multilayer stack).

5) Mirror profile errors and microroughness act independently from each other, so we can disentangle the mirror figure error HEW term from the scattering term as follows:

$$HEW^2(\lambda) \approx H_0^2 + H^2(\lambda), \qquad (7)$$

where $H_0$ represents the HEW due to geometrical deformation, whereas the energy-dependent term $H(\lambda)$ denotes the HEW caused by the XRS.

In the following we shall see how the $H(\lambda)$ function is analytically related to the surface PSD $P(f)$.

### 3.1 From the PSD to the HEW

The $H(\lambda)$ function for an optical system with $N = 1,2,...$ identical reflections can be computed from the PSD $P(f)$ along with the integral equation[24]

$$\int_{f_0}^{2/\lambda} P(f)df = \frac{\lambda^2}{16\pi^2 \sin^2 \vartheta_i} \ln\left(\frac{2N}{2N-1}\right), \qquad (8)$$

that allows one to calculate the lower integration limit $f_0$. Mirror shells with Wolter-I profile fall in the case $N = 2$. Then, for all $N$, $H(\lambda)$ is calculated along with the Eq. 4 in this approximate form

$$H(\lambda) = \frac{2\lambda f_0}{\sin \vartheta_i}. \qquad (9)$$

Application of these formulae to all photon wavelength of interest returns the desired function. The HEW trend for the mirror shell is then obtained by using the Eq. 7. We shall provide in Sect. 4 a detailed example of HEW computation from a PSD, using the Eqs. 8 and 9.

It is worth noticing that the Eq. 8 is derived supposing that the X-ray detector that collects the scattered photons is very large. This assumption justifies the upper integration limit $2\cos\theta_i/\lambda \approx 2/\lambda$, that corresponds to a photon back-scattering, even if the PSD at so large frequencies usually is usually irrelevant for the calculation, because it has so small values that contributes negligibly to the integral in Eq. 8. However, in practice the detector size has always a finite angular radius $r$ and scattered photons beyond $r$ are lost, so they do not contribute to the imaging degradation (they cause, indeed, effective area loss). If the HEW is computed from a EE function being normalized to its maximum measured value, the upper integration limit in the Eq. 8 should be modified as follows:

$$\frac{2}{\lambda} \rightarrow f_r = \frac{r \sin \vartheta_i}{\lambda}. \qquad (10)$$

The effect of this substitution is to diminish the HEW because the high-frequency tail of the PSD is ruled out from the integration, so one should integrate down to smaller $f_0$ to return the constant on right-hand side of Eq. 8.

### 3.2 From the HEW to the PSD

The Eqs. 8 and 9 can be inverted in order to derive the PSD from a $H(\lambda)$ function. We can immediately derive the frequency $f_0$ at which the PSD is evaluated from the Eq. 9:

$$f_0(\lambda) = H(\lambda) \frac{\sin \vartheta_i}{2\lambda}, \qquad (11)$$

whilst the PSD at the frequency $f_0$ at can be immediately computed[24] from the derivative of the ratio $H(\lambda)/\lambda$:

$$\frac{P(f_0)}{\lambda} \frac{d}{d\lambda}\left(\frac{H(\lambda)}{\lambda}\right) + \frac{1}{4\pi^2 \sin^3 \vartheta_i} \ln\left(\frac{2N}{2N-1}\right) \approx 0. \qquad (12)$$

Therefore, if one wants to find the surface PSD responsible for to a measured HEW trend in a definite photon wavelength band, this equation can be very useful. Similarly, it can be used to translate a HEW requirement into a surface PSD, that can be assumed as roughness tolerance for the X-ray mirrors. For the Eq. 12 to be valid, we have to suppose that the HEW has been computed from the PSF over an angular range much larger than the HEW. For instance, if the X-rays range of interest is the interval [$\lambda_M$, $\lambda_m$] (with $\lambda_M > \lambda_m$) and the corresponding $H(\lambda)$ values increase (monotonically) in the range [$H_m$, $H_M$], with $H_M > H_m$, the PSD can be computed up to the maximum spatial frequency

$$f_{MAX} = H_M \frac{\sin \vartheta_i}{2\lambda_m};$$ (13)

for the validity of Eq. 12, it has to be $f_r >> f_{MAX}$.

It is important to note that, even if the Eq. 12 is able to provide with a PSD in a definite spectral range (determined by the extent of the X-ray wavelength range and the HEW values), it does not return alone a complete surface characterization equivalent to the requested/measured HEW trend, because the PSD can be non-unique beyond $f_{MAX}$. Nevertheless, the Eq. 8 with $\lambda = \lambda_m$ and $f_0 = f_{MAX}$ sets another constraint on the PSD at larger frequencies than $f_{MAX}$: That relation is able to bound the PSD at high frequencies, that influences all the HEW trend, and in particular the one at higher photon energies. As before, whenever the HEW becomes comparable to the detector size, the substitution $2/\lambda \rightarrow f_r$ should be done (Eq. 10).

### 3.3 The case of a fractal surface

If the power-law model (Eq. 3) is suitable to extensively describe the surface PSD, it can be used to derive an analytical expression for the $H(\lambda)$ function[24], by means of the Eq. 8. The most interesting result is that the HEW *inherits the power-law trend from the PSD*,

$$H(\lambda) \propto \left( \frac{\sin \vartheta_i}{\lambda} \right)^{\gamma},$$ (14)

and the spectral index $\gamma$ is related to that of the PSD, $n$, along with the simple, algebraic relation

$$\gamma = \frac{3-n}{n-1}.$$ (15)

This result is interesting because the requirement *1 < n < 3* required by the theory of fractal surfaces becomes equivalent to the statement $\gamma > 0$, indicating that the HEW is an increasing function of the photon energy $E \propto 1/\lambda$. Moreover, it highlights the relevant impact that the spectral index *n* has on the $\lambda$ dependence. For example, if $n \rightarrow 3$, $\gamma \rightarrow 0$: the HEW becomes nearly energy-independent if the spectral index *n* goes close to its maximum allowed value. On the contrary, the case $n \rightarrow 1$ implies $\gamma \rightarrow +\infty$ and the increase of the HEW becomes very quick. Finally, a $n = 2$ power-law index would cause a linear growth of the $H(\lambda)$ term ($\gamma = 1$). This discussion makes apparent the importance of a steeply-decreasing surface PSD in X-ray mirror design and fabrication, especially in the hard (> 10 keV) X-ray band, where $H(\lambda)$ can dominate the figure error contribution.

In general, the Eq. 14 is *not* valid for PSDs that are *not* power-laws. Instead, the most general relation between the *local* power-law index of the PSD,

$$-n_{f_0} = \frac{d(\log P)}{d(\log f_0)},$$ (16)

and the corresponding *local* power-law index of the $H(\lambda)$ function (the correspondence is established by the Eq. 9),

$$-\gamma_\lambda = \frac{d(\log H)}{d(\log \lambda)},$$ (17)

is given by the more complex equation

$$n_{f_0} = \frac{3+\gamma_\lambda}{1+\gamma_\lambda} - \frac{\lambda}{(1+\gamma_\lambda)^2} \frac{d\gamma_\lambda}{d\lambda},$$ (18)

that reduces to the Eq. 15 for constant $\gamma_\lambda$ (i.e. for a power-law PSD). However, if the HEW has only a finite number of power-law index changes, the derivative in Eq. 18 is zero almost everywhere. In other words, the Eq. 14 retains its validity also *locally*, except at frequencies where power-law indexes change. The derivation of the Eq. 18 is postponed in appendix A.

# 4. COMPARISON OF THE TWO METHODS FOR THE SIMBOL-X OPTICAL MODULE

A comparison of the results of the simulation described in Sect. 2 and the results that can be obtained using the method described in Sect. 3 has been performed. This will also allow us not only to cross-check the two different approaches, but also to return an estimation of the X-ray scattering impact on the HEW of the SIMBOL-X optical module. This will also enable the evaluation of tolerable surface roughness levels.

The PSD taken in order to compute the HEW trend using the results of Sect. 3 is a simulation at low frequency, aimed at returning a slowly-increasing HEW from 15 arcsec at 1 keV up to 20 arcsec at 30 keV. If we assume the HEW at 1 keV to be essentially due to figure errors, adopting a HEW interpolation of the two values and using Eqs. 8,13,14 (assuming N=2 and $\theta_i$ = 0.18 deg, the incidence angle of an intermediate mirror shell of SIMBOL-X), we can derive a PSD from the HEW trend. The PSD is plotted in Fig. 4 (marks) and covers only the low-frequencies regime (spatial wavelengths $\ell = 1/f > 408$ μm).

As mentioned in Sect. 3, the HEW requirement does not allow us to derive the PSD at higher frequencies, unless we extend the required HEW trend at higher energies. However, in order to assign realistic values to the PSD, we adopt for the high-frequency regime a measured PSD. The sample under test, assumed to be representative for the achievable smoothness levels in X-ray optics, is a Gold-coated X-ray mirror shell prototype sample, available at INAF/OAB. Metrological instrumentation operated at INAF/OAB (AFM, optical profilometers, XRS measurements at 8.05 keV) allowed us to characterize its roughness in terms of PSD at spatial wavelengths shorter than 600 μm. The merging of the simulated and measured PSDs (Fig. 4, solid line) returns a roughness description over a very wide spectral range. In addition, it represents a realistic surface finishing level at all spatial wavelengths considered here.

The HEW scattering term was thereby derived from the PSD in Fig. 4 via Eqs. 8 and 9. The computation was done from 1 to 65 keV, for each SIMBOL-X mirror shell incidence angle. In order to make the results comparable with the simulation carried out in Sect. 2, we accounted for the finite size of the detector ($r$ = 310 arcsec) by modifying the upper integration limit (Eq. 10). Moreover, 15 arcsec were added in quadrature (Eq. 7) to the scattering term $H(\lambda)$ to allow for mirror figure errors: the same contribution was assumed in the simulations in Sect. 2. As an example, calculated HEW($E$) for three SIMBOL-X mirror shells are plotted in Fig. 5. Notice that the HEW trends exhibit an initial saturation, followed by a steep divergence at the highest energies. This can be ascribed to the *slope change* of the PSD at ~200 μm (see Fig. 4), that causes a sudden increase of the local spectral exponent of the HEW, $\gamma_\lambda$ (Eq. 18), when λ becomes sufficiently small to set the frequency $f_0$ in the "smoother" part of the PSD.

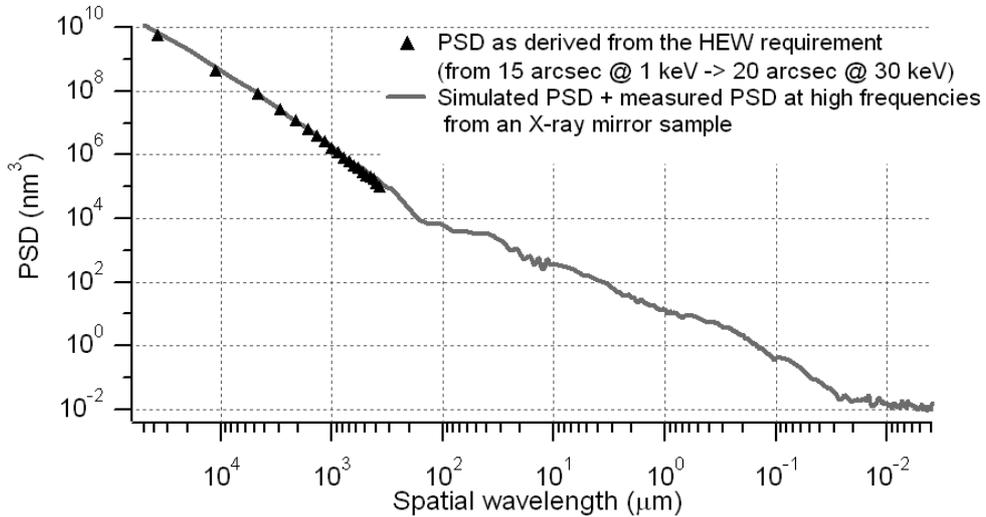

Fig. 4: the PSD being utilized to simulate the HEW of the SIMBOL-X optical module. The PSD at low frequencies (triangles) is derived from the requirement that the HEW increases slowly from 15 arcsec at 1 keV up to 20 arcsec @ 30 keV. The remaining high-frequency PSD is measured from a Gold-coated X-ray mirror.

Now, the $HEW_T$ of the entire optical module can be *approximately* calculated by averaging the HEW of each mirror shell, $HEW_k$, over the respective effective areas $A_k(E)$, $k = 1\ldots100$:

$$HEW_T(E) = \frac{\sum_{k=1}^{100} HEW_k(E) A_k(E)}{\sum_{k=1}^{100} A_k(E)}. \qquad (19)$$

The approximate validity of this assumption is also proven in another paper of this volume[26]. Notice that the effect of the average over the effective areas is a partial compensation of the HEW divergence, because the mirror shells with the largest incidence angles (that enhance the XRS) have also the lowest energy cut-off in the reflectivity. Therefore, the contribution to the HEW of the largest shells at high energies is relatively modest. It is worth noting *that the calculated HEW is 15 arcsec at 1 keV and equals 22 arcsec at 30 keV, close to the required 20 arcsec for SIMBOL-X at that energy.*

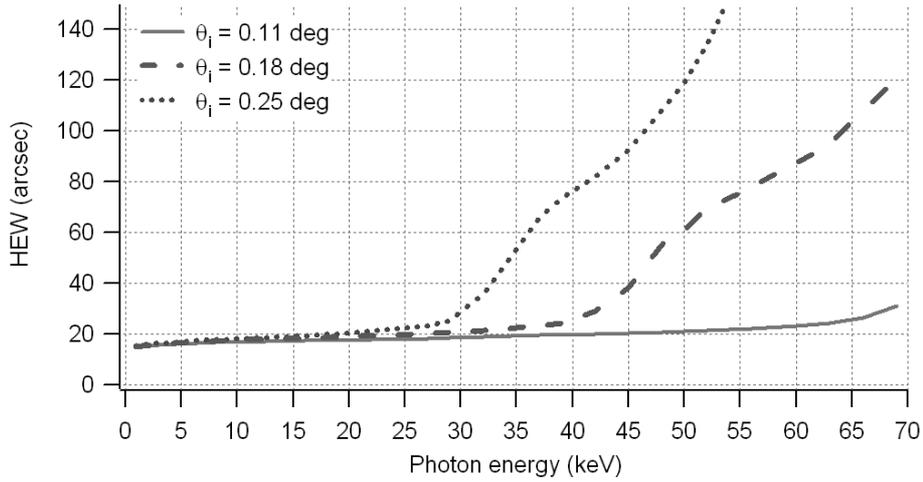

Fig. 5: the HEW as computed from the PSD, for three mirror shells of SIMBOL-X at the incidence angles of: 0.11 deg (the smallest one) 0.18 deg (the 55$^{th}$ shell) and 0.25 deg (the largest one). The HEW increases with the energy and the increase is more marked at the largest angles. The oscillations in the HEW are caused by small fluctuations of the PSD.

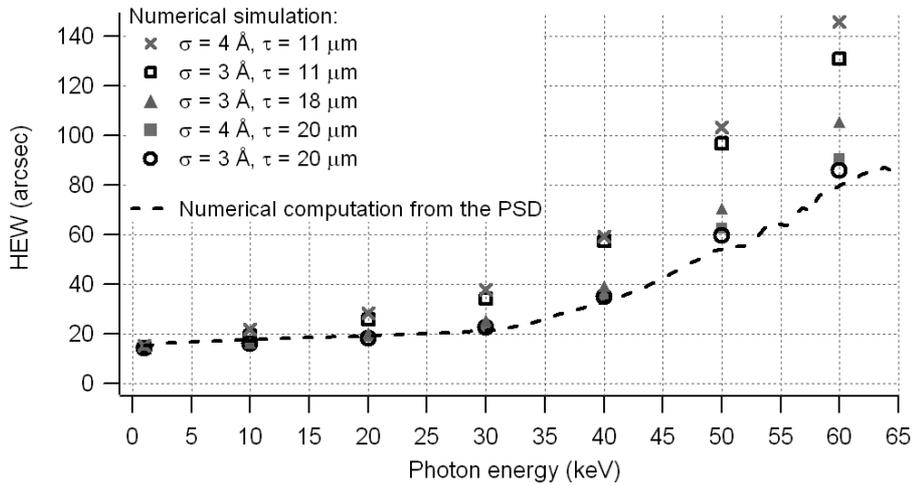

Fig. 6: comparison of the HEW trend as derived via the modified ray-tracing routine (marks) for several couples rms/correlation length, and the analytical computation from the PSD. The agreement is satisfactory if we assume $\sigma = 3$ Å (or less) and $\tau = 20$ μm as input parameters for the numerical simulation.

The result of the calculation is overplotted (see Fig. 6) to the results of the modified ray-racing routine referred to in Sect. 2: the HEW(*E*) function just computed is *consistent with the numerical simulation with the parameter values σ = 3 Å (or less) and τ = 20 μm*. Now, to complete the benchmarking of the two procedures, we have to verify the consistency of these parameter values with the PSD we started from.

The mentioned consistency check can – in principle – be achieved by simply applying the Eqs. 2 and 5. However, one should notice that *the definitions of these parameters depend strongly on the frequency window $\Delta f$*. Thus, we should find out the spectral band that determines the HEW at each photon energy/incidence angle, before computing the rms and the correlation length from the PSD. To do this, notice that only a *limited set of scattering angles $[\theta_s^m, \theta_s^M]$ contribute to the HEW*[24]. This range of scattering angles, for fixed values of $\lambda$ and $\theta_i$, *defines a spatial frequency window $[f_m, f_M]$*, along with Eq. 4.

More precisely, the spectral range of interest for each $\lambda$ would in principle equal the integration range of Eq. 8: however, as we assumed a limited size of the detector over which the PSF is observed (310 arcsec radius), the upper integration limit has to be modified according to Eq. 10. Therefore, we may adopt

$$\Delta f = \left( \frac{H \sin \vartheta_i}{2\lambda}, \frac{r \sin \vartheta_i}{\lambda} \right) \tag{20}$$

as frequency band of interest where we should calculate the parameters $\sigma$ and $\tau$.

Unfortunately, the evaluation is made more complicate by the dependence of the spectral band $\Delta f$ on $\theta_i$ and $\lambda$, whereas the numerical simulation was performed for single-valued σ and τ, and for the entire set of angles of the SIMBOL-X optical module. We can, indeed, estimate the *effective* values for the parameters σ and τ, by calculating the average spectral band $<\Delta f>$ over all $\lambda$, and over all the incidence angle set of SIMBOL-X mirror shells. The resulting spectral band turns out to be (0.02 μm$^{-1}$, 0.07 μm$^{-1}$). Integrating the PSD (Eqs. 2 and 5) in Fig. 4 within these limits, we can easily compute the effective parameters, σ and τ. The results are *σ = 2.8 Å and τ = 29 μm*, vs. 3.0 Å and 20 μm which were required by the numerical simulation. Therefore, the two methods return *similar* values for the parameters characterizing the surface roughness, even if not exactly the same ones. It should be noted that the remaining discrepancy could be ascribed to the very approximate method we adopted for the comparison: a more detailed analysis would have required to simulate the photon distribution at each photon energy and for each mirror shell, using the actual σ - τ parameters, as computed in the respective spatial frequencies band (Eq. 20).

## 5. CONCLUSIONS AND FINAL REMARKS

In the previous sections we have discussed some possible approaches to the problem of the HEW determination for an X-ray optic, in a photon energy band, from the roughness characterization of the mirrors surface. To this aim, we utilized a modified ray-tracing routine, allowing the simulation of the photons spread on the focal plane due to mirror deformations and surface roughness from two roughness parameters; this method takes the advantage of a simultaneous treatment of profile errors and roughness. We also utilized an alternative, analytical approach to directly derive the HEW scattering term from the surface PSD, which can be applied to any PSD and takes the advantage of providing with analytical solutions to translate a PSD into a HEW trend, and vice versa. We performed this way a HEW calculation for the SIMBOL-X optical module, and compared the results of the two methods, finding a quite satisfactory agreement.

The methodologies discussed above can be therefore applied to the definition of microroughness tolerances for future X-ray telescopes. An example has already been discussed in Sect. 4 for the case of the SIMBOL-X telescope, leading to the conclusion that the proposed PSD (or the proposed rms and the correlation length), assuming $H_0 = 15$ arcsec, *returns a HEW of 22 arcsec at 30 keV, near the 20 arcsec requirement*. This small discrepancy can be explained as follows: the required 20 arcsec at 30 keV was used to derive the low-frequency PSD via Eqs. 11 and 12. The PSD at higher frequency than $f_{MAX} = (408 \text{ μm})^{-1}$ could not be inferred, because our requested HEW was defined in a too limited range of energies to do that. Indeed, we are able to set a constraint on the integral of the PSD (Eq. 8) from $f_{MAX}$ (Eq. 13), up to the $f_r$ (Eq. 10). In practice, always assuming for the incidence angles the average value $\theta_i = 0.18$ deg, we can derive for the integral of the mirrors PSD, between $f_{MAX} = (408 \text{ μm})^{-1}$ and $f_r = (7 \text{ μm})^{-1}$, the constraint

$$\int_{f_{MAX}}^{f_r} P(f) df \leq 5.6 \text{ Å}, \tag{21}$$

whereas the numerical integration of the PSD in the mentioned spectral range returns 6.3 Å. Therefore, the proposed PSD fulfills the smoothness requirements only approximately.

Detailed calculations of the XRS term of the HEW for the future X-ray telescopes SIMBOL-X, EDGE, XEUS, and the consequent formulation of microroughness PSD requirements, are carried out in another paper of the present proceedings [26]. In that paper, it is also shown the relevant impact of high-frequency microroughness in hard X-rays: in order to preserve the imaging quality in hard X-rays, future developments of the X-ray optics have to concentrate on the microroughness damping, in particular at spatial wavelengths shorter than a few hundreds micron.

## APPENDIX A. GENERAL RELATION BETWEEN PSD AND HEW POWER LAW INDEXES

Using the Eq. 11, one can easily derive an equivalent form of the Eq. 12:

$$-\frac{df_0}{d\lambda}P(f_0) = \frac{\lambda}{8\pi^2 \sin^2 \vartheta_i} \ln\left(\frac{2N}{2N-1}\right). \quad (22)$$

The Eq. 17, in turn, can be developed as follows:

$$-\gamma_\lambda = \frac{\lambda}{H}\frac{dH}{d\lambda} = \frac{1}{f_0}\frac{d(\lambda f_0)}{d\lambda} = 1 + \frac{\lambda}{f_0}\frac{df_0}{d\lambda}, \quad (23)$$

whereas the Eq. 16 can be rewritten, using the Eq. 22, as

$$-n_{f_0} = f_0\frac{d(\log P)}{df_0} = f_0\frac{d}{df_0}\left[\log \lambda - \log\left(-\frac{df_0}{d\lambda}\right)\right], \quad (24)$$

note that all the constants that appear in the Eq. 22 have been dropped out, because their logarithm is also a constant and would be canceled by the derivation in the Eq. 24. Furthermore, notice that $df_0/d\lambda < 0$ (see Eq. 11) as long as the HEW is a decreasing function of $\lambda$, so the second logarithm in the Eq. 24 has a positive argument. Now we have, from Eq. 23,

$$-\frac{df_0}{d\lambda} = \frac{f_0}{\lambda}(1+\gamma_\lambda): \quad (25)$$

the substitution in the Eq. 24 yields

$$-n_{f_0} = f_0\frac{d}{df_0}\left[2\log \lambda - \log f_0 - \log(1+\gamma_\lambda)\right]. \quad (26)$$

Carrying out the derivations and using the Eq. 25, we obtain

$$n_{f_0} = -2\frac{f_0}{\lambda}\frac{d\lambda}{df_0} + 1 + f_0\frac{d\log(1+\gamma_\lambda)}{df_0} = \frac{3+\gamma_\lambda}{1+\gamma_\lambda} + \frac{d\log(1+\gamma_\lambda)}{d(\log f_0)}: \quad (27)$$

this formula provides the relation between the two spectral indexes we were looking for, without hypotheses on the PSD shape. After some handling of the derivative and using again the Eq. 25, it takes the suggestive form

$$n_{f_0} = \frac{3+\gamma_\lambda}{1+\gamma_\lambda} - \frac{\lambda}{(1+\gamma_\lambda)^2}\frac{d\gamma_\lambda}{d\lambda}. \quad (28)$$

## ACKNOWLEDGMENTS

We thank O. Citterio, S. Basso, P. Conconi, F. Mazzoleni *(INAF-OAB, Merate, LC, Italy)*, S. L. O'Dell, B. Ramsey *(NASA-MSFC, Huntsville, AL, USA)*, P. Gorenstein, S. Romaine *(SAO- CfA, Boston, MA, USA)* for useful discussions. This research is funded by ASI (Italian Space Agency), *MUR* (the Italian Ministry for Universities), *INAF* (the National Institute for Astrophysics).